\documentclass[aps,epsf,11pt]{revtex4-1}

\def\a0{$a_0$}

\begin{document}

Essay written for the Gravity Research Foundation 2018 Awards for Essays on Gravitation.

\title{Does GW170817 falsify MOND?}

\author{R.H. Sanders}
\affiliation{Kapteyn Astronomical Institute,
P.O.~Box 800,  9700 AV Groningen, The Netherlands;
email: sanders@astro.rug.nl}
\date{\today}
\begin{abstract}
  The gravitational-wave
   event GW170817 and the near-simultaneous corresponding gamma-ray burst (GRB 170817 A)
   falsify modified gravity theories in which the gravitational geometry differs non-conformally from
   physical geometry.  Thus, the observations of this event definitively  rule out
   theories, such as TeVeS, a suggested relativistic
   extension of Milgrom's modified Newtonian dynamics (MOND), that predict  a 
   significantly different Shapiro delay for electromagnetic
   and gravitational radiation.  While not 
   falsifying MOND per se, GW170817 severely constrains relativistic extensions of
   MOND to theories that do not rely on additional matter-coupling fields but
   rather upon modified field equations for one universal gravitational and physical metric. 
   Here I mention a simple preferred-frame theory as an example. 
\end{abstract}
\maketitle
Modified Newtonian dynamics (MOND) is an alternative to dark matter
proposed by Milgrom 35 years ago \cite{milg83}.  The basic idea is that the
modification of dynamics or gravity occurs below a critical acceleration,
$a_0 \approx cH_0/6$, that appears to have a cosmological significance.
In its original manifestation, MOND was essentially an algorithm
permitting the calculation of the effective gravitational 
acceleration in an astronomical object
from the observed distribution of baryonic matter.  And it works on the scale of
galaxies as judged by the successful prediction of the observed form and amplitude
of galaxy rotation curves with, in a number of cases, no free parameters, apart from
the  new universal constant $a_0$ \cite{bf}.  
The fact that such an algorithm exists is challenging for non-interacting
dark matter that does not naturally permit such a parameter-free prediction.
However, the absence of a credible relativistic 
extension of MOND has frustrated considerations of
cosmology and cosmogony;    MOND in original form is clearly incomplete.

The first step toward a more fundamental theory was made by Bekenstein and Milgrom \cite{bekmil}
who wrote down a non-relativistic Langrangian from which a modified
Poisson equation may be derived:
$$\nabla\cdot{[\mu(|\nabla{\Phi}|/{a_0}}) \nabla{\Phi}] = 4\pi G\rho.\eqno(1)$$
Here $\mu$ is an unspecified function that must have the asymptotic limits
$\mu(x) = 1$ where $x>>1$ (the Newtonian limit) and $\mu(x) = x$
where $x<<1$ (the modified gravity limit).

A relativistic extension as a modified
scalar-tensor theory is immediately suggested.  To general relativity one adds
a scalar field with a non-standard Lagrangian;  i.e., 
$$L = {\sqrt{-g}\over {16\pi}}[R-{{a_0}^2} F(\xi)], \eqno(2)$$ where $F$ is a function of
$\xi$, the standard scalar field invariant in terms
of  $a_0$, i.e. $\xi={{\Phi_{,\alpha}\Phi^{,\alpha}}/ {{a_0}^2}}.$
In the modified Poisson equation $\mu = dF/d\xi$ and
can be chosen such that the theory becomes a weakly coupled Brans-Dicke
theory where $\nabla\Phi >>a_0$; i.e., $\mu \rightarrow \omega>>10000$ with $\omega$ as 
the Brans-Dicke parameter.  In the
opposite limit, the force about a point mass $M$ becomes 
$\nabla\Phi = \sqrt{GMa_0}/r$ exceeding the Newton-Einstein force, $\nabla\phi$, 
at accelerations below $a_0$.

Consistency with the Equivalence Principle in its weak form (the universality of free fall) 
is typically guaranteed in scalar-tensor modifications by assuming that  
matter couples to the scalar field jointly with the gravitational or Einstein metric (${g}_{\mu\nu}$) 
via a conformal factor to form a new metric -- a ``physical metric" $\tilde{g}_{\mu\nu} = f^2(\Phi)g_{\mu\nu}$.
Particles follow geodesics of the physical metric, and as usual
the paths of photons and other relativistic particles trace null geodesics.
But it is trivial to see that
null geodesics of the two conformally related metrics coincide.  In other words, photons
are not affected by the scalar field, 
and this has an immediate observational consequence if
an anomalous force is to replace dark matter.  Photons, in contrast to the
slowly moving constituents of an astronomical object, do not respond to
the putative dark matter but only to the baryonic content of the object.  Then in
gravitational lensing of a distant source by an intervening massive structure
(a cluster of galaxies, for example), the effective lensing mass should be significantly
smaller than the traditional Newtonian dynamical mass -- in stark contradiction to
the observations \cite{bs}.

The lensing contradiction led Bekenstein to consider a more general relation
between physical and gravitational geometry:  the so-called disformal
transformation.  Basically, a conformal transformation takes the geometry
described a metric and stretches it in a manner that is isotropic but space-time dependent.
A disformal transformation, on the other hand, picks a preferred direction for
additional stretching.  Because one wishes space to be 
isotropic, that direction is generally
taken to be the time coordinate in some preferred frame -- which is to say, at a fundamental level 
the theory violates Lorentz invariance of gravitational phenomena (not particle dynamics).  

The mechanism for choosing this special direction 
is taken to be a normalized vector field, $A^\mu$, that is dynamical but points in the positive 
time direction of the preferred frame.  
This is the basis of Bekenstein's Tensor-Vector-Scalar theory or TeVeS \cite{bek04}
where the relation between the physical and gravitational metrics is given by 
$$\tilde{g}_{\mu\nu} = \exp({-2\Phi})g_{\mu\nu}  -2\sinh{(2\Phi)} A_\mu A_\nu. \eqno(3)$$
For such a transformation 
null geodesics of the physical metric no longer coincide with null geodesics
of the gravitational metric;  the photons respond to the scalar field (or dark matter) and
the relationship between the deflection of photons due to the total weak-field
force is identical to that provided by general relativity with dark matter. 

{\it But} while photons track null geodesics of the physical metric (and therefore
``feel" the dark matter), gravitational waves follow null geodesics of the 
Einstein metric and are not affected by the scalar field or putative dark matter.
Given a source of the gravitational radiation within a galaxy, such as an inspiraling
and coalescing binary neutron star, the potential well 
from which the gravitational waves emerge is that due only to the baryonic content of the
galaxy and thus significantly more shallow than that experienced
by the emerging photons that see the baryonic galaxy plus scalar field (``dark halo").  
Thus the Shapiro delay for photons from any such event is typically greater than that of the 
gravitational radiation by several hundred days \cite{boretal}. 

As we all know such a burst of
gravitational radiation, GW170817, has been detected by
LIGO/Virgo;  the frequency and observed flux is consistent with a coalescing pair of binary neutron 
stars at a distance of about 40 Mpc.   A corresponding 
gamma-ray burst was detected by the Fermi satellite in the direction of an early-type galaxy also at a distance
of 40 Mpc; the gamma-ray burst followed the  gravitational-radiation event by less than 
2 seconds \cite{ligo} -- inconsistent (by a factor of at least $10^7$) with that expected if the gravitational and 
electromagnetic radiation track null geodesics of two distinct disformally related metrics. 
The implication is that $\tilde{g}_{\mu\nu} \equiv g_{\mu\nu}.$ 
{\it Gravitational geometry corresponds to the physical geometry of space-time to high
precision}  (as Einstein assumed).
So while the observation does not falsify MOND as a non-relativistic theory, it seriously constrains 
relativistic extensions of MOND.  The theory cannot rely upon
adding an additional field that couples to matter disformally with the Einstein
metric, as TeVeS, but rather upon a modification of the field equation 
itself and hence of the Einstein metric.

There are several existent theories 
that might fit this bill \cite{defw,milg14}, but here I focus on one in particular because of its simplicity:  
that is, a modification
connected to a preferred timelike foliation of space-time provided by a dynamical
scalar field, the ``Khronon", that does not directly couple to matter \cite{BPS}.  The unit normal
to this foliation defines a vector field that may contribute various 
terms to the field Lagrangian -- terms that are quadratic in first derivatives of the vector 
(an Einstein-Aether theory \cite{tj,zfs}).  But the
unique aspect here is that only one of  these -- the term becoming
the three-dimensional gravitational acceleration in the preferred frame --  is
included and assumed to vanish at high gravitational accelerations ($>>a_0$) 
thus restoring general relativity in this limit \cite{blanmar,rsand}.

Following Blanchet and Marsat, the field Lagrangian of the theory is given by
$$L = {\sqrt{-g}\over {16\pi}}[R-{{a_0}^2} F(\chi)] \eqno(4)$$ where matter couples directly to
the gravitational metric as in general relativity.  This appears to be identical to the Lagrangian in
the relativistic version of  the
original Bekenstein-Milgrom theory (eq.\ 2), but in fact $\chi$ is formed by the
ordinary three-dimensional acceleration
in the preferred frame,  i.e., $\chi = \phi^{,i}\phi_ {,i/}{a_0}^2$
where $\phi$ is the first Newtonian
potential ($g_{00} = -1-2\phi$) in the weak-field expansion; it is not an additional scalar field but part of the
gravitational metric.  Therefore, the theory modifies the Einstein equation and the
gravitational metric.

Taking the weak field to first
order one finds that the two Newtonian potentials, $\phi$ and $\psi$ ($g_{ij} = \delta_{ij}(1-2\psi)$),
are the same.  This 
provides gravitational lensing equal to that of general relativity;  there
is no need to construct a physical metric disformally related to the
gravitational metric. We also find the Bekenstein-Milgrom modified Poisson equation (eq.\ 1),
but now with $\mu = 1+{a_0}^2 dF/d{\chi}.$
The form of this function can be chosen to achieve MOND phenomenology where 
$\nabla\phi < a_0$ but general relativity at high accelerations
(including a cosmological constant on the natural order of ${a_0}^2$)

This may or may not be the correct theory (several issues are swept under the
carpet in the brief description given here), but it does
suggest a qualitative scenario.  Observationally there is clearly
a preferred frame -- the frame at rest with respect to the cosmic microwave background 
radiation.  We do not detect
any dynamical effects of this preferred frame locally because at the 
high accelerations
prevailing in the Solar System, the world becomes Lorentz-invariant to high precision
and described by general relativity.  In the outer parts of galaxies, the transition between
local general relativity and preferred-frame cosmology,  the phenomenology
of MOND appears -- a phenomenology commonly ascribed to dark matter.
\acknowledgements
I have had many discussions on this subject with 
Jacob Bekenstein over three decades and
benefited greatly from his deep and intuitive understanding of relativity.  
I also gratefully acknowledge Phillip Helbig for a critical reading of this manuscript.

\end{document}